\documentclass[iop, apj]{emulateapj}
\usepackage{threeparttable}
\usepackage{multirow}
\usepackage{times}
\usepackage{enumitem}
\usepackage{url} 
\usepackage{color}
\usepackage{amsmath,bm}
\usepackage{verbatim}

\shorttitle{}
\shortauthors{ X.-Y. Li et al.}

\begin{document}

\title{Mapping the Galactic disk with the LAMOST and Gaia Red clump sample:\\ IV: the kinematic signature of the Galactic warp}

\author{ X.-Y. Li\altaffilmark{1}}
\author{ Y. Huang\altaffilmark{1,4}}
\author{ B.-Q. Chen\altaffilmark{1}}
\author{ H.-F. Wang\altaffilmark{1,3}}
\author{ W.-X. Sun\altaffilmark{1}}
\author{ H.-L. Guo\altaffilmark{1}}
\author{ Q.-Z. Li\altaffilmark{2}}
\author{ X.-W. Liu\altaffilmark{1,4}}

\altaffiltext{1}{South-Western Institute for Astronomy Research, Yunnan University, Kunming 650500, People's Republic of China; {\it yanghuang@ynu.edu.cn {\rm (YH)}; x.liu@ynu.edu.cn {\rm (XWL)}}}
\altaffiltext{2}{Yunnan Observatories, Chinese Academy of Sciences, Kunming, Yunnan 650011, People's Republic of China}
\altaffiltext{3}{LAMOST Fellow}
\altaffiltext{4}{Corresponding authors}

\begin{abstract}
Using a sample of nearly  140,000 red clump stars selected from the LAMOST  and Gaia Galactic surveys, we have mapped mean vertical velocity $\overline{V_{z}}$ in the $X$--$Y$ plane for a large volume of the Galactic disk (6 $< R < 16$\,kpc; $-20 <\phi<50^{\circ}$ ; $|Z| < 1$\,kpc). A clear signature where $\overline{V_{z}}$ increases with $R$ is detected for the chemically thin disk. The signature for the thick disk is however not significant, in line with the hot nature of this disk component. For the thin disk, the warp signature shows significant variations in both radial and azimuthal directions, in excellent agreement with the previous results of star counts. Fitting the two-dimensional distribution of $\overline{V_z}$ with a simple long-lived static warp model yields a line-of-node angle for this kinematic warp of about $12.5^{\circ}$, again consistent with the previous results.

\end{abstract}

\section{Introduction}
Disk warping in the outer regions of spiral galaxies ($>50\%$) is a very common phenomenon (e.g. Saha et al. 2009). In general, the inner disk of a spiral galaxy is largely flat whereas the outskirts show significant warp signature. The warp amplitude increases strongly with radius, reaching as large as a few times of the inner disk scale height. Being a typical spiral galaxy, the Milky Way (hereafter MW) also shows clear warp in the outer disk. The Galactic warp was first detected by Kerr (1957) with H~{\sc i} 21-cm line observation. This was confirmed later by Weaver \& Williams (1974) and Henderson (1979). Not only the neutral gas, other components of the Galactic disk also show that the Galactic outer disk is strongly warped, including the stars (Efremov et al. 1981; Reed 1996; L{\'o}pez-Corredoira et al. 2002b), the molecular clouds (Wouterloot et al. 1990) and the interstellar dust grains \citep{2006A&A...453..635M,2019MNRAS.483.4277C}. Studies show that one part of the Galactic outer disk bends up from the Galactic plane to the north Galactic pole, whereas the other part bends down. Further studies indicate that the warp amplitude not only increases strongly with radius but also changes with azmithual angle. The line-of-node angle with respect to the Sun-Galactic Centre line is estimated to range between $-$5 and  26$^{\circ}$ by different groups using different tracers (e.g. L{\'o}pez-Corredoira et al. 2002b; Momany et al. 2006; Chen et al. 2019b; Skowron et al. 2019).

Theoretically, warping of a spiral galaxy is generally interpreted as the response of the disk to perturbations. Specially, for our MW, the perturbations may come from: i) the interactions of the Galactic disk with nearby satellite galaxies (e.g. the Large and Small Magellanic Clouds or the Sagittarius dwarf galaxy; Weinberg 1995; Garc{\'{\i}}a-Ruiz et al. 2002; Bailin 2003); ii) effects of the triaxial dark-matter halo \citep{1988MNRAS.234..873S,1999ApJ...513L.107D}; or iii) the accretion of infalling intergalactic gas \citep{1999MNRAS.303L...7J,2002A&A...386..169L,2006MNRAS.365..555S}. While many scenarios have been proposed, the exact origin of the Galactic warp remains unclear. Further information of the kinematic signature of the Galactic warp would be invaluable to clarify the situation.

Prior to the first Gaia data release, several studies (e.g. Miyamoto et al. 1988; L{\'o}pez-Corredoira et al. 2014) have attempted to unravel the kinematic signature of the Galactic warp, using catalogs of ground-based proper motion measurements. The results are inconclusive due to the limited accuracy of proper motions employed. With the release of Gaia DR1 \citep{2016A&A...595A...1G,2016A&A...595A...4L}, accurate measurements of proper motions and parallaxes for over two million stars become available. With the data, Poggio et al. (2017) have detected signature of kinematic warp by using nearby OB stars. With the Gaia DR1 and the spectroscopic information from the RAVE and LAMOST surveys, Sch{\"o}nrich \& Dehnen (2018) and Huang et al. (2018) have calculated values of vertical velocity $V_{z}$, azimuthal velocity $V_{\phi}$  and vertical angular momentum $L_{z}$ for stars in the Solar neighborhood. They find that mean vertical velocity $\overline{V_{z}}$  increases with $V_{\phi}$, $L_{z}$ and guiding center radius $R_{g}$. The trends are consistent with the predictions of long-lived Galactic warp model. Recently, the Gaia DR2 has been released, providing accurate parallax and proper motion measurements of about 1.3 billion stars. With the new data, accurate kinematics of the Galactic disk has been mapped by Gaia Collaboration et al. (2018c). Poggio et al. (2018; hereafter P18) use two samples, one of stars of the upper main sequence and another of red giant stars, and study the kinematic signature of the Galactic warp in the $X$--$Y$ plane out to a distance of 7\,kpc from the Sun. However, for their giant sample, only 24 per cent of the stars have line-of-sight velocities. The distances, estimated from the Gaia parallaxes, for more distant stars may also suffer from serious systematics (e.g. Sch{\"o}nrich et al. 2019). Recently, Huang et al. (2020; hereafter Paper I), based on data from the LAMOST and Gaia surveys, have published a sample of about 140,000 red clump (RC) stars with accurate measurements of distance, proper motions and stellar atmospheric  parameters (effective temperature $T_{\rm eff}$, surface gravity log\,$g$ and metallicity [Fe/H]), line-of-sight velocity $V_{\rm los}$ and $\alpha$-element to iron abundance ratio [$\alpha$/Fe]. The sample allows one to study the warp signature over a large volume for both the chemically thin and thick populations.
 
The paper is organized as follows. In Section 2, we define the coordinate systems and describe the data used. The results are presented and discussed in Section 3. Finally, a summary is presented in Section 4.

%figure1
\begin{figure*}[!htbp]
\begin{minipage}[t]{0.5\linewidth}
\centering
\includegraphics[width=7.in]{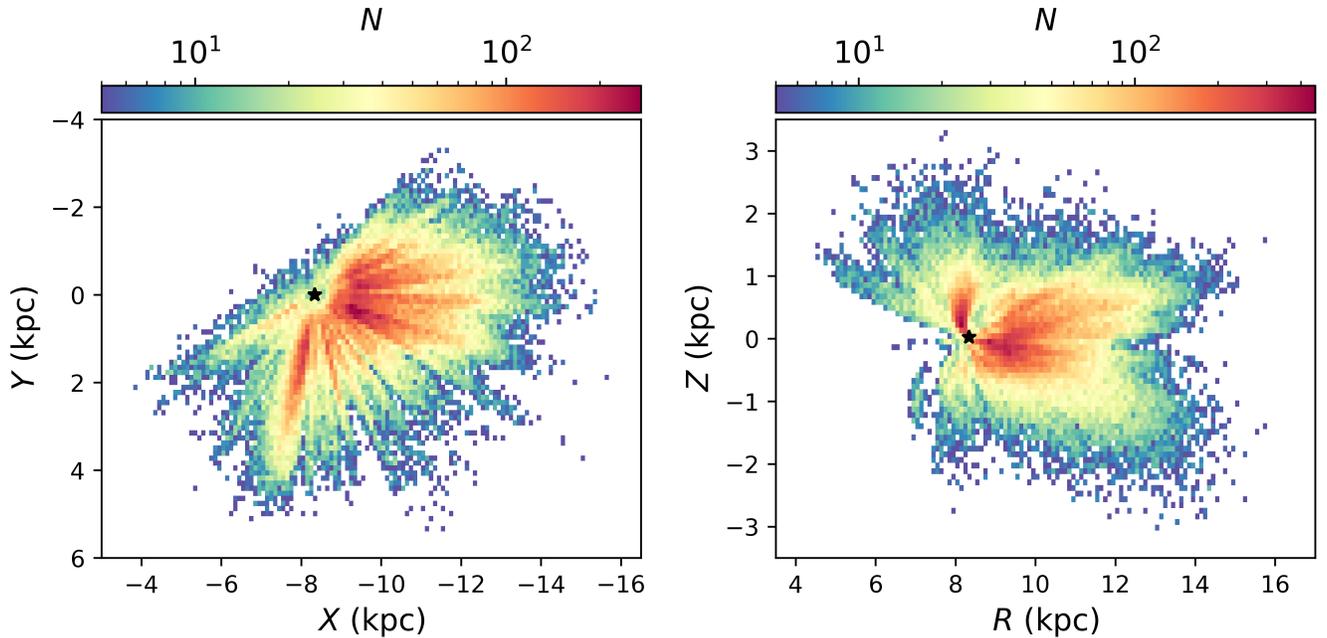}
\end{minipage}
\caption{Spatial distribution of the RC sample stars in the $X$--$Y$ (left) and  $R$--$Z$ (right) planes. The stellar number densities (in bins of size 0.1\,kpc for both axes) are represented by colorbars on top, with no less than 5 stars in each bin. The Sun is represented by a black star at $X = -8.34$\,kpc, $Y = 0$\,kpc and $Z=0.025$\,kpc.}
\end{figure*}

%figure2
\begin{figure*}[!htbp]
\begin{minipage}[t]{0.5\linewidth}
\centering
\includegraphics[width=3.5in]{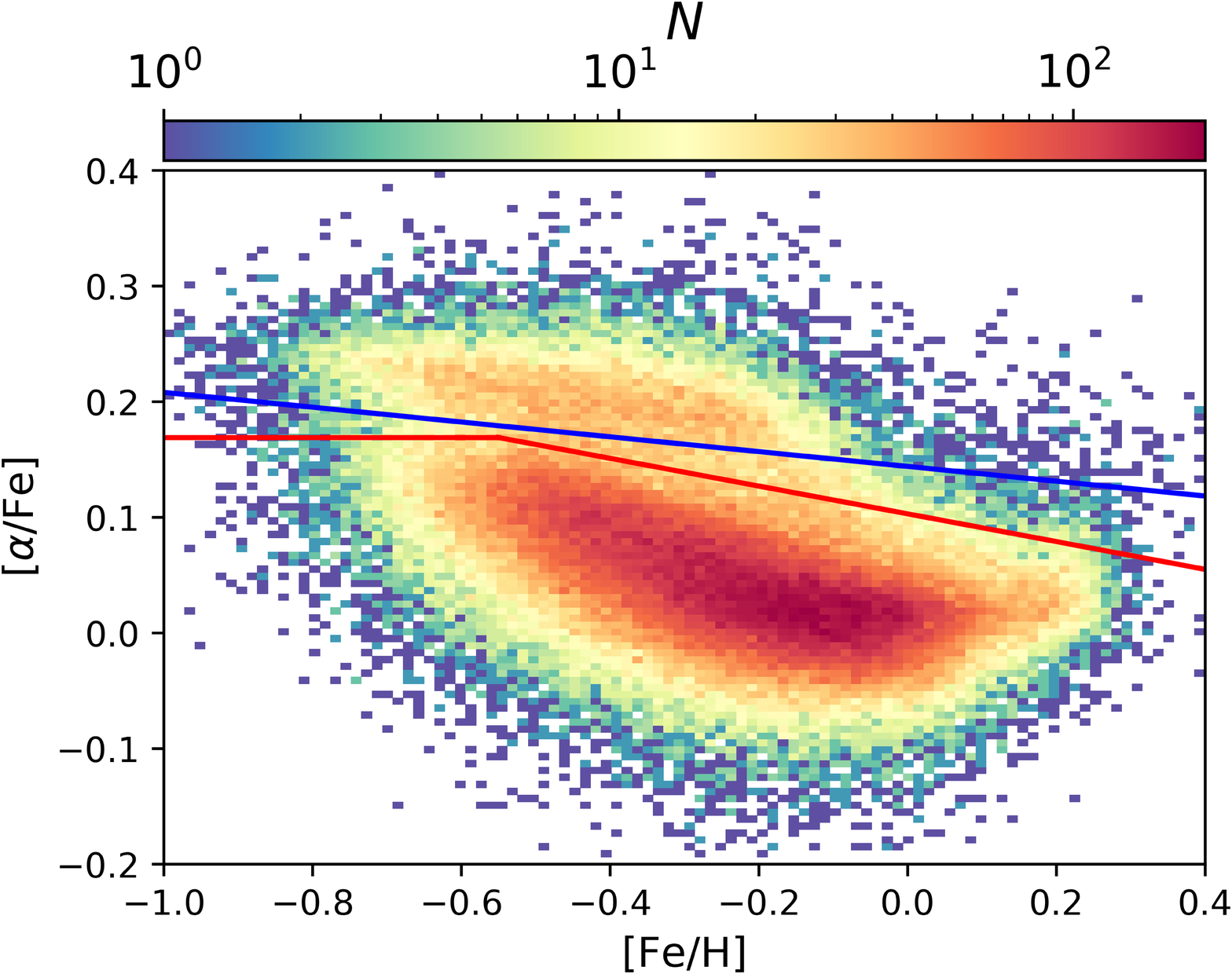}
\end{minipage}
\begin{minipage}[t]{0.5\linewidth}
\centering
\includegraphics[width=3.5in]{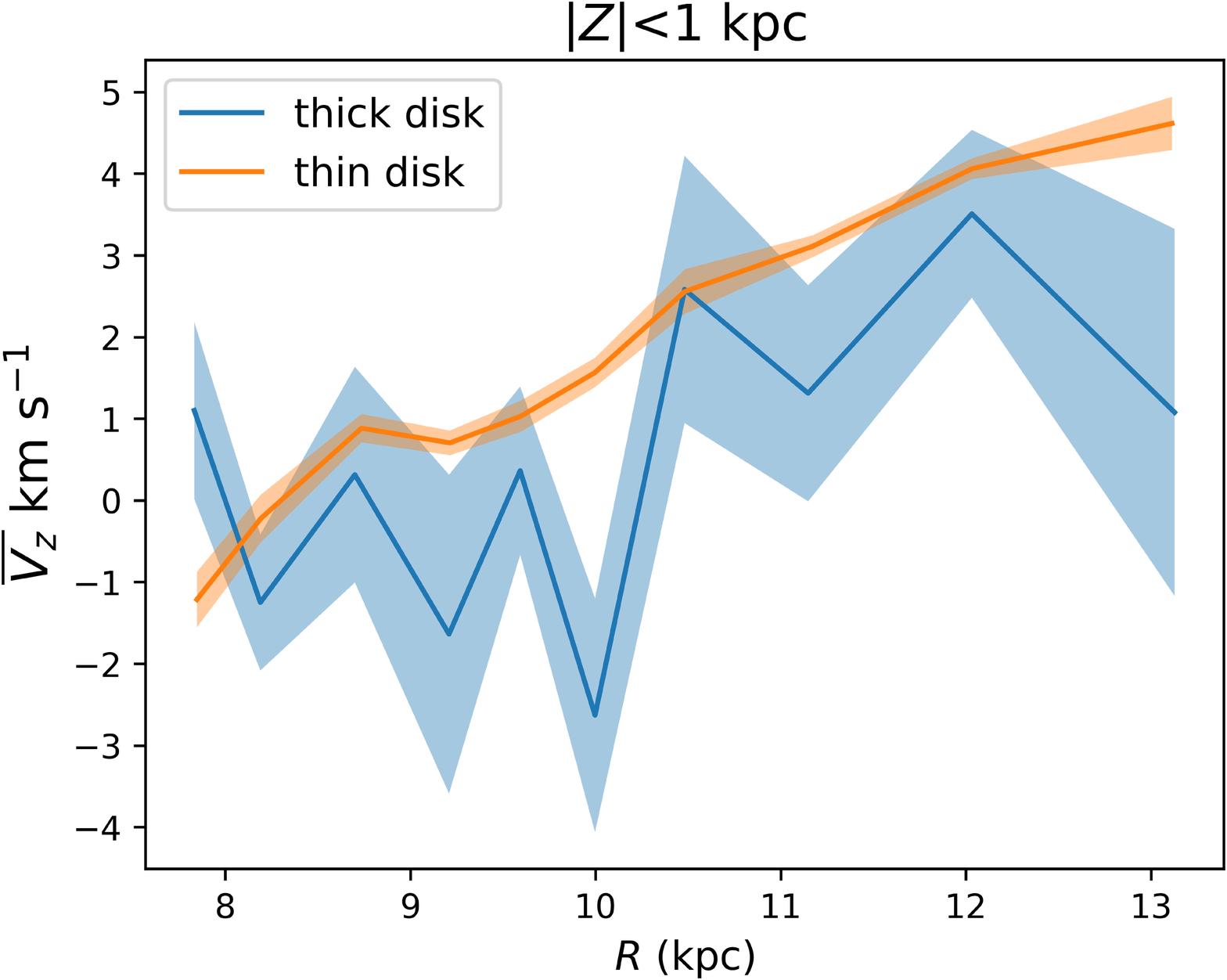}
\end{minipage}
\caption{{\it Left panel:} Distribution of the RC sample in the  [Fe/H]--[$\alpha$/Fe] plane. Thick disk stars lie above the blue line while those of the thin disk fall below the red line. The stellar number densities in bins of size 0.014 dex in horizontal axis and of 0.006 dex in vertical axis are indicated by the top colorbar. {\it Right panel:} The blue and orange solid lines represent mean vertical velocities of the chemically thin and thick stars, respectively. The binsize in $R$ is 0.4\,kpc for $R < 10.2$\,kpc and 1\,kpc for $R \ge 10.2$\,kpc, comparable with the typical distance uncertainties. The shaded areas represent the $\pm1\sigma$ uncertainties (estimated with a bootstrapping procedure) of the mean vertical velocities.}
\end{figure*}

%figure3
\begin{figure}[!htbp]
\begin{minipage}[t]{0.5\linewidth}
\centering
\includegraphics[width=3.5in]{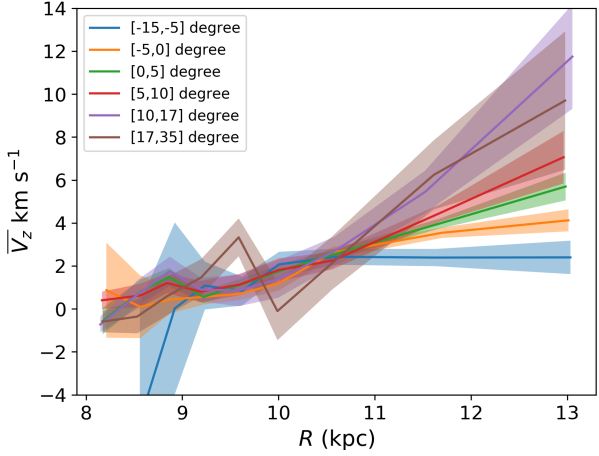}
\end{minipage}
\caption{The R--$\overline{V_{z}}$ diagram of 94,028 thin disk sample stars. The blue, orange, green, red, purple and magenta solid lines represent the mean vertical velocities of stars in azimuthal angle ranges $\phi$ $\in$ [-15.0, -5.0], [-5.0, 0.0], [0.0, 5.0], [5.0, 10.0], [10.0, 17.0] and [17.0, 35.0]  deg, respectively, with the shaded areas representing the $\pm1\sigma$ uncertainties (estimated with a bootstrapping procedure) of the mean vertical velocities.}
\end{figure}

%figure4
\begin{figure*}
\begin{center}
\includegraphics[width=6.in]{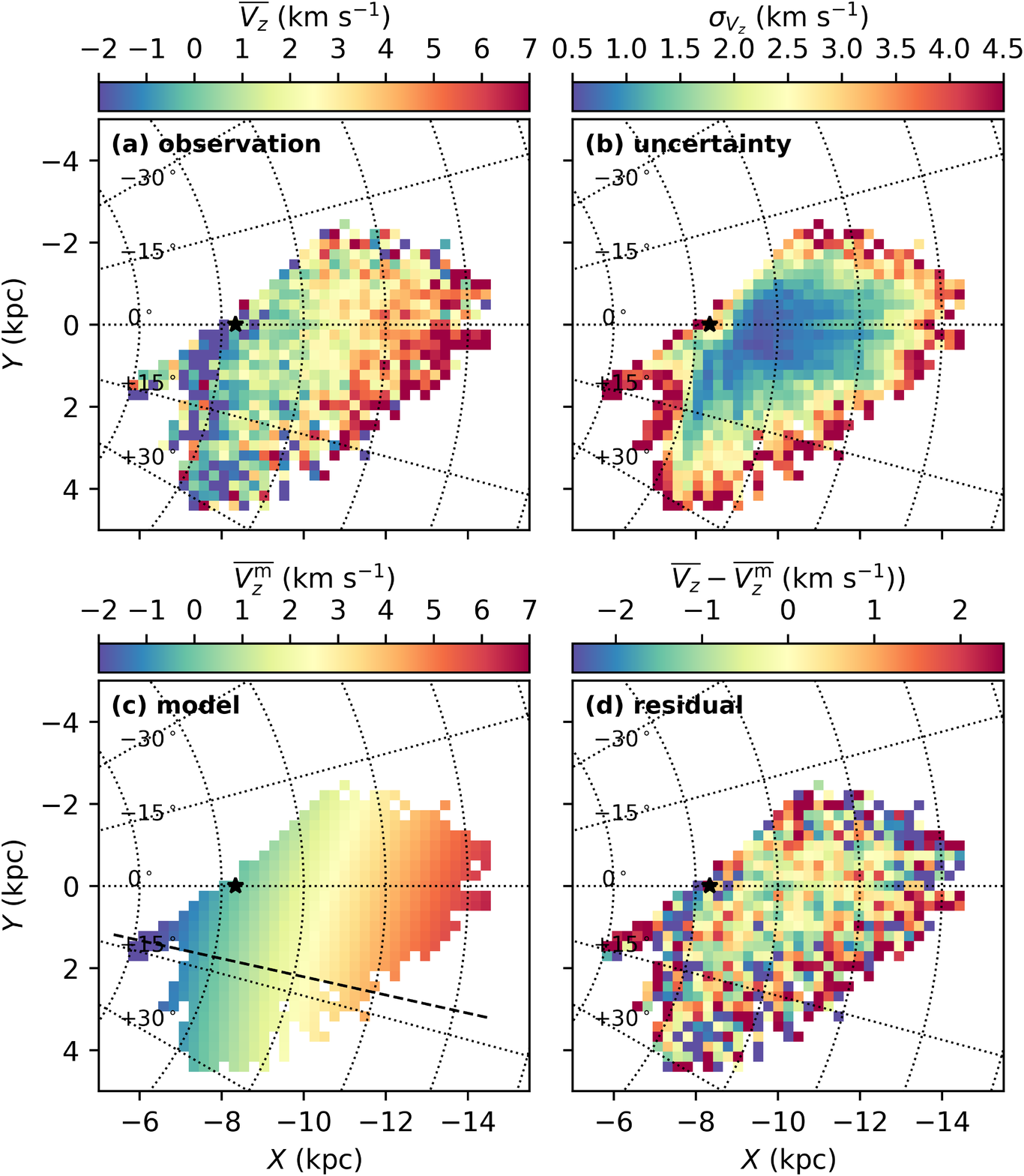}
\caption{Distributions of mean vertical velocities in the $X$-$Y$ plane. The panel (a) shows the result from our observational data for the chemically thin disk stars, with no less than 15 stars in each bin. The panel (b) shows the uncertainties $\sigma_{V_z}$ (estimated with a bootstrapping procedure) of the mean vertical velocities. The panel (c) shows the result of our best-fit model. The position of the resulted line-of-node is marked by a black dashed line. The panel (d) shows the fit residuals (observation minus model). The azimuthal angle increases clockwise. The Sun is represented by a black star at $X = -8.34$\,kpc and $Y = 0$\,kpc.  The binsize in $X$ and $Y$ axes is 0.25 kpc.}
\end{center}
\end{figure*}

%figure5
\begin{figure*}[!htbp]
\begin{minipage}[t]{0.5\linewidth}
\centering
\includegraphics[width=3.5in]{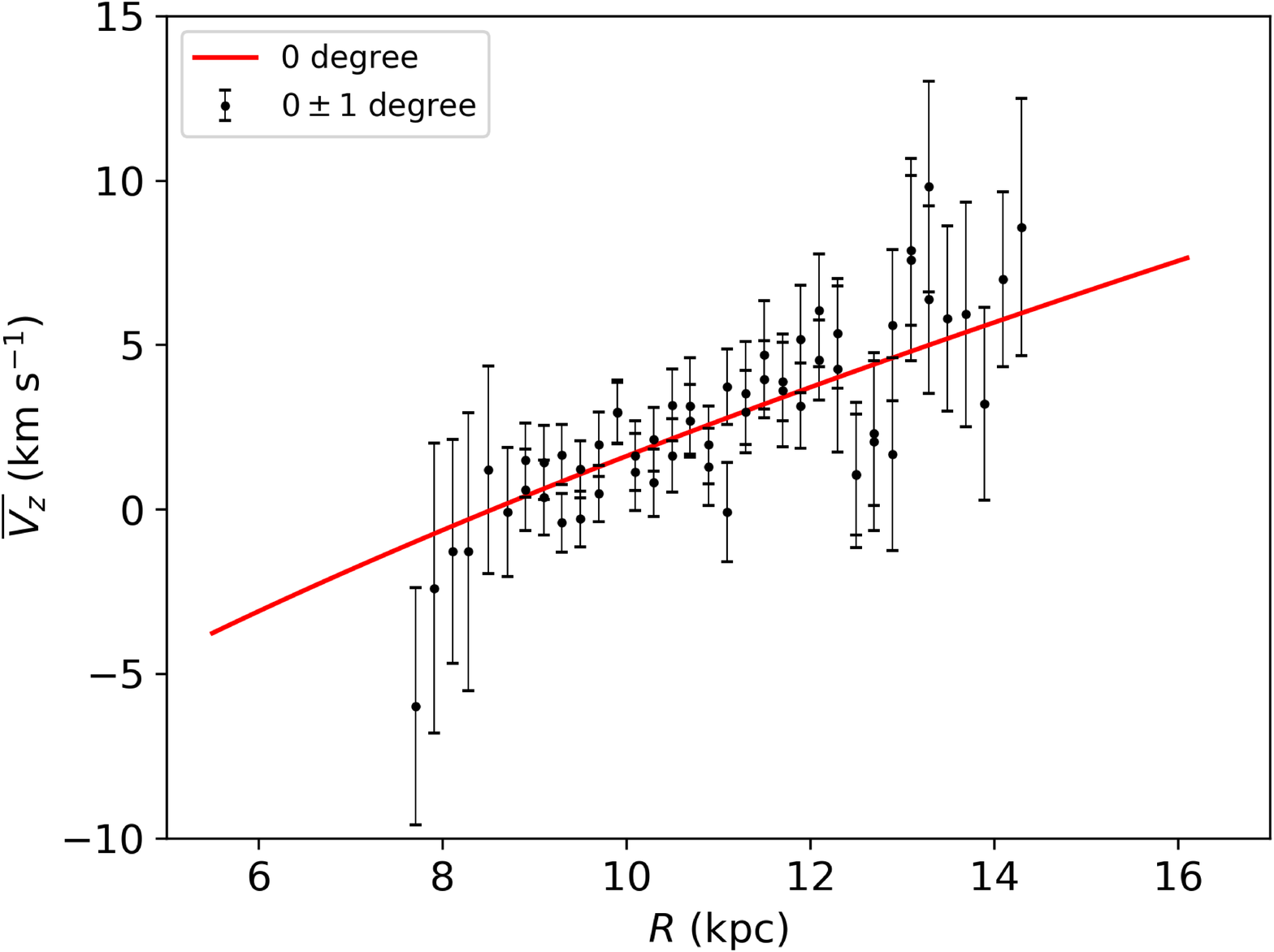}
\end{minipage}
\begin{minipage}[t]{0.5\linewidth}
\centering
\includegraphics[width=3.5in]{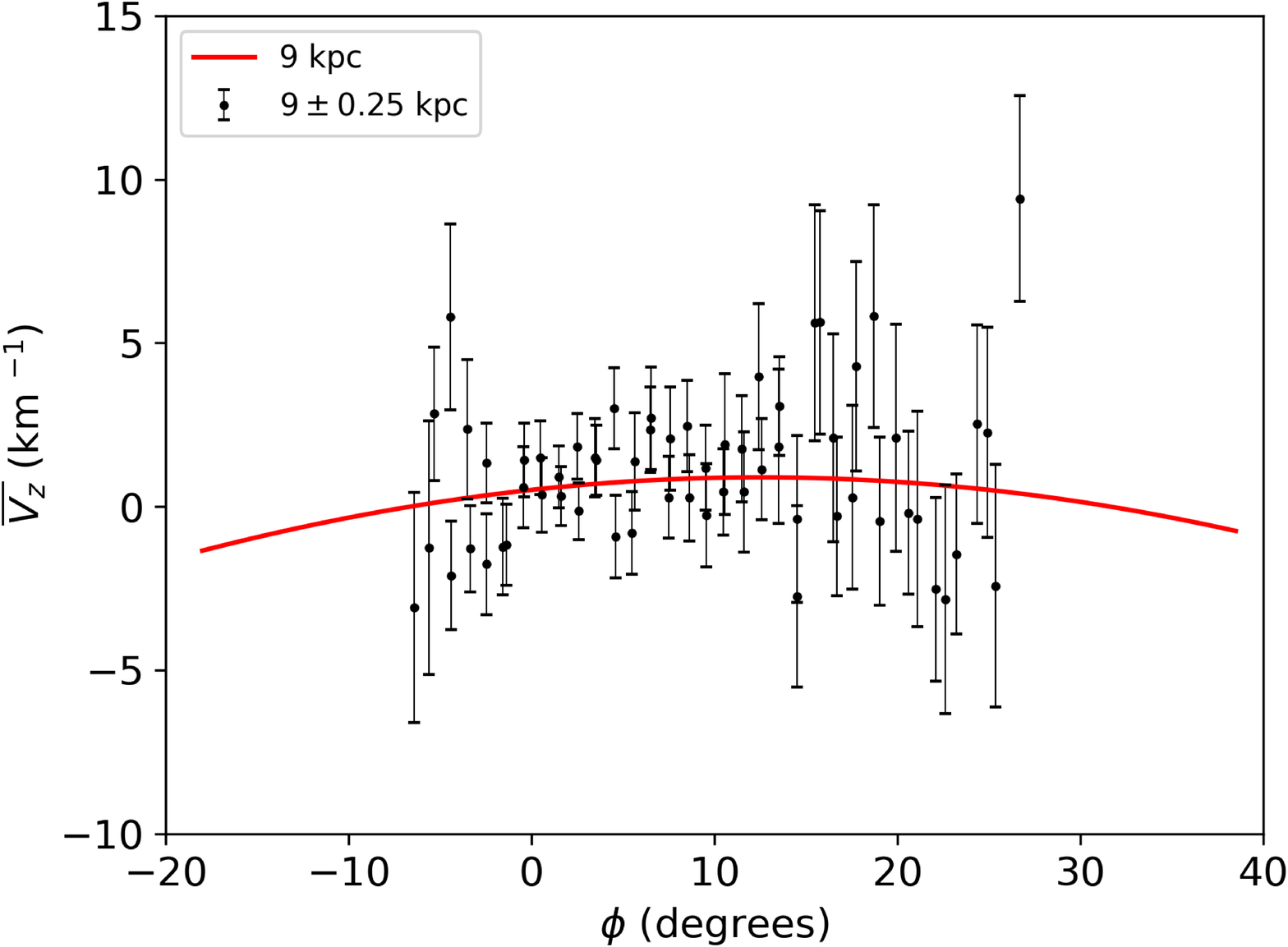}
\end{minipage}
\caption{{\it Left panel:} Mean vertical velocity as a function of $R$ for $\phi = 0^{\circ} \pm 1^{\circ}$. {\it Right panel:} Mean vertical velocity as a function of $\phi$ for $R = 9\pm 0.25$\,kpc. The red line in both panels represents the best-fit (see Section\,3.2). The calculations of the mean velocity and its uncertainties are given in Section\,3.2.}
\end{figure*}

\section{coordinate systems and data}
\subsection{Coordinate systems}
In this paper, we use two coordinate systems. A Galactocentric cylindrical coordinate system ($R$,\,$\phi$,\,$z$), along with associated velocity components (\,$V_{R}$,\,$V_{\phi}$,\,$V_{z}$), is defined with $R$ the projected Galactocentric distance, $\phi$ the azimuthal angle increasing in the direction of Galactic rotation and $z$ the height above the Galactic plane in the direction of the north Galactic pole. The velocity components are calculated from the sky positions, distances, line-of-sight velocities and proper motions, using the standard transformations from Johnson \& Soderblom (1987). We adopt Galactocentric distance of the Sun $R_{0}$ of 8.34 kpc \citep{2014ApJ...783..130R} and circular velocity at the Solar radius of $V_{c} (R_{0})=238$\,km\,s$^{-1}$ \citep{2016MNRAS.463.2623H}. We take Solar motions with respect to the Local Standard of Rest ($U_{\odot}$,\,$V_{\odot}$,\,$W_{\odot}$)\,$=$\,$(11.10,\,12.24,\,7.25)$\,km\,s$^{-1}$ \citep{2010MNRAS.403.1829S}. Other values of the Solar motions (e.g. Huang et al. 2015) are also tried and the results obtained are essentially the same. Also used is a right-handed Cartesian Galactocentric coordinate system ($X$,\,$Y$,\,$Z$), with $X$ pointing towards the Galactic center, $Y$ in the direction of Galactic rotation and $Z$ towards the north Galactic pole. 

\subsection{Data}
LAMOST is a 4-metre quasi-meridian reflecting Schmidt telescope equipped with 4000 fibers distributed in a field of view of about 20 sq.deg. It can simultaneously collect spectra per exposure of upto 4000 objects, covering the wavelength range 3800--9000\,\AA \ at a resolving power $R$ of about 1800 \citep{2012RAA....12.1197C}. The five-year Phase-I LAMOST Regular Surveys started in the fall of 2012 and completed in the summer of 2017. The scientific motivations and target selections of the surveys are described in detail in Deng et al. (2012), Zhao et al. (2012) and Liu et al. (2014). Atmospheric parameters ($T_{\rm eff}$, log\,$g$, [Fe/H]), line-of-sight velocity $V_{\rm los}$ and $\alpha$-element to iron abundance ratio [$\alpha$/Fe] of the targeted stars are derived with the LAMOST Stellar Parameter Pipeline at Peking University (LSP3; Xiang et al. 2015, 2017). The second data release of Gaia have been made available to the community since April 2018, providing accurate parallax and proper motion measurements for about 1.3 billion stars (Gaia Collaboration et al. 2018a). Typical uncertainties of the parallaxes are 0.04 mas for bright sources ($G< 14$\,mag), 0.1 mas  at $G = 17$\,mag and 0.7 mas at $G = 20$\,mag. For the proper motions, typical uncertainties are 0.05, 0.2 and 1.2 mas \,yr$^{-1}$ at $G < 14$\,mag, $G = 17$\,mag and $G = 20$\,mag, respectively.

In the current work, a sample of nearly 140,000 RC stars has been used. The sample is described in Paper I, constructed with data from the LAMOST and Gaia surveys. Given the standard candle nature of RCs, distances of those stars have been measured with a typical accuracy of 5-10 per cent, preciser even than values yielded by the Gaia parallax measurements for stars beyond 3-4\,kpc. With the derived distances, line-of-sight velocities, proper motions, [Fe/H] and [$\alpha$/Fe] values for the LAMOST and Gaia RC sample stars, we have derived 3D positions and velocities for all the sample stars, and examine the velocity field of disk stars. The current work concentrates on the vertical velocity field of disk stars of different populations in a large disk volume. The spatial distribution of our sample stars is presented in Fig.\,1. The sample covers a large volume of the Galactic disk of $-16 \leq X \leq -4$\,kpc, $-3 \leq Y \leq 6$\,kpc and $|Z| \leq 3$\,kpc.

\section{Results and discussion}

The mean vertical velocity field of the (outer) Galactic disk can be significantly perturbed in the long-lived warp model (e.g. Drimmel et al. 2000). With the current RC sample, we explore how the mean vertical velocity field varies with $R$ and in the disk plane (i.e. $X$-$Y$ plane) for the different stellar populations. 

\subsection{The kinematic warp of the Galactic disks}
Before mapping the mean vertical velocity field, we first exclude sample stars with vertical velocity uncertainties $e_{V_{z}}$ (estimated for the individual stars with a Monte Carlo method) larger than 15\,km\,s$^{-1}$ and $|V_{z} - \overline{V_{z}}| > 3\sigma_{z}$. The latter cut is used to remove significant outliers in the vertical velocity distribution of our sample. The distribution of the remaining 133,061 stars in the [Fe/H]-[$\alpha$/Fe] plane is presented in the left panel of Fig.\,2. As the plot shows, a bimodal distribution is clearly seen. As in the previous studies (e.g. Bensby et al. 2005; Lee et al. 2011;  Haywood et al. 2013), we define cuts to separate the two populations, one of the chemically thin disk and another of the thick disk in the plane. The cuts result in 94,028 and 5,212 chemically thin and thick disk stars within $|Z| < 1$\,kpc (close to the Galactic plane), respectively. The mean values of vertical velocity $\overline{V_z}$ as a function of $R$ for the chemically thin and thick disk stars are presented in the right panel of Fig.\,2. The plot shows that $\overline{V_z}$ of the chemically thin disk stars increases with $R$, from $-1.5$\,km\,s$^{-1}$ at $R \sim 8$\,kpc to $4.5$\,km\,s$^{-1}$ at $R \sim 13$\,kpc. The trend is similar to that reported in P18 (see their Fig.\,3). 

In addition to what found for the chemically thin disk population, we have tried to detect the  warp signature for the chemically thick disk population. Fig.\,2 shows that $\overline{V_z}$ of the chemically thick stars present a very weak positive trend with $R$, from $-1$\,km\,s$^{-1}$ at $R = 8$-10\,kpc to $2.5$\,km\,s$^{-1}$ at $R = 12$-13\,kpc. Due to the limited number of thick disk stars, the uncertainties of mean vertical velocities are large (the typical uncertainty is about $1.6$\,km\,s$^{-1}$). According to above analysis, only 2.2$\sigma$ detection is found for the thick disk population on the kinematic warp. In the future, more thick disk stars are required to reduce the random errors of mean $V_z$ to clarify whether there is a clear kinematic warp for the thick disk population. On the other hand, a weak/insignificant kinematic warp signature for the thick disk population is in line with the hot nature of orbits of the thick disk stars (i.e. of large velocity dispersions in all directions; e.g. Chiba \& Beers 2000; Bensby et al. 2003; Parker et al. 2004). Because of their hot nature, the thick disk stars are less sensitive to the warp perturbations than the thin disk stars. Secondly, the large velocity dispersions (especially in the radial direction) of the thick disk population can smooth the warp signature along $R$ direction. To fully understand the weak/insignificant warp signature of the thick disk population, both observational efforts (by obtaining more thick disk stars) and theoretical dynamical modeling are required. 

For the thin disk population, we extract the kinematic warp signature shown in Fig.\,2 in different azimuthal slices, from $-15$ to $35$ deg. The results are presented in Fig.\,3. The positive trend of $\overline{V_z}$ increasing with $R$, i.e. the kinematic warp signature, is seen in almost all the azimuthal slices. The amplitude of warp signature increases with azimuthal angle $\phi$ and reaches a maximum in slice $\phi \in$\,[10, 17] deg, and then decreases slightly in the last azimuthal slice. In the scenario of a long-lived static Galactic warp model, the variations of warp amplitude with $\phi$ presented here indicate an angle of line-of-node between 10 and 17 deg. To obtain a preciser estimate, we fit the distribution of $\overline{V_z}$ in the $X$-$Y$ plane with a naive long-lived static warp model in the next Subsection.

\subsection{Fitting the $\overline{V_z}$ distribution with a simple model}

To fit the kinematic warp with a theoretical model, the distribution of mean vertical velocity in the Galactic plane is presented in the panel (a) of Fig.\,4. The general trend of variations with $R$ and $\phi$ are similar to those found in P18. Theoretically, the Galactic warp can be simulated with either by a transient model (e.g. Christodoulou et al. 1993; Debattista \& Sellwood 1999) or a long-lived one (e.g. Smart et al. 1998; L{\'o}pez-Corredoira et al. 2002b; Poggio et al. 2017). Observationally, the results from star counting prefer the later (e.g. Djorgovski \& Sosin 1989; L{\'o}pez-Corredoira et al. 2002b). To fit the distribution of $\overline{V_z}$ found here for the chemically thin disk population, we have adopted the long-lived static warp model proposed by Poggio et al. [2017; see their Eq.\,(3)]. We have simply assumed that mean azimuthal velocity $V_{\phi}$ is constant, and $R_{\omega}$ is zero (i.e. the  warp starts at the Galactic center; e.g.  L{\'o}pez-Corredoira et al. 2002b). We have also add a constant $c$, corresponding to mean vertical velocity at $R=0$ or $|\phi-\phi_{0}|=\frac{\pi}{2}$ to the warp equation. The final simplified model adopted here is then,
\begin{equation}
\overline{V_{z}} = aR^{b}\cos(\phi-\phi_{0})+c,
\end{equation}
where $\phi_0$ is angle of line-of-node of the warp.

To fit the observational data, values of $\overline{V_z}$ are calculated for the individual bins of $R$ and $\phi$, the uncertainties of the mean vertical velocities are estimated with a bootstrapping procedure. The binsize in $R$ is 0.2\,kpc and that in $\phi$ is allowed to vary but no less than 1$^{\circ}$ such that there are at least 40 stars in a bin. This results in total of 776 bins. Values of $\overline{V_z}$ of those bins are then fitted with the model described by Eq.\,(1), using a MCMC method. The best-fit model yields parameters,  

\begin{align}
  a &= 3.97^{+1.77}_{-1.24},\\
  b &= 0.64^{+0.10}_{-0.09},\\
  c &= -15.31^{+2.40}_{-2.97} \rm \,km \ s^{-1} and\\
  \phi_{0} &= 12.5^{+2.0}_{-1.8} \rm \,degree.
\end{align}

The angle of line-of-node obtained above agrees very well with recent estimates using Pulsars (Yusifov 2004), RCs and red giants (Momany et al. 2006) and Cepheids (Chen et al. 2019b) as tracers. We note that this is the first estimate of angle of line-of-node of the Galactic warp using kinematic data. However, the other best-fit value of the warp amplitude $a$ found here is not consistent with the results reported by previous star count analysis (e.g., L{\'o}pez-Corredoira et al. 2002b; Yusifov 2004; Momany et al. 2006; Chen et al. 2019b). This may indicate more additional kinematic parameters are required for explaining this kinematic warp (like a precession; e.g. Poggio et al. 2020). Also, we note that the very large negative value for the parameter $c$ is unconvincing. It's probably a consequence of assuming that the warp starts at the center of the Galaxy. In Fig.\,5, we show two typical examples of the best-fit for constant $\phi = 0^\circ$ and for constant $R = 9$\,kpc. Generally, the model fits the observational data quite well. The model predicted $\overline{V_z}$ distribution, the uncertainties of the mean vertical velocities and the fit residuals are also present in Fig.\,4. The residuals are largely within 1\,km\,s$^{-1}$.

The variations of angle of line-of-node as a function of $R$ is explored by Chen et al. (2019b). Using over one thousand classical Cepheid stars, they find that the angle of line-of-node first decreases with $R$ for $R$ between 8 and 12\,kpc and then increases with $R$ for $R$ between 12 and 15\,kpc, and tends to twist near $R = 15.5$\,kpc. They claim that the increase of the angle of line-of-node between 12 and 15\,kpc is evidence that the warp in the outer disk is predominately induced by torques associated with the massive inner disk. Our current data of mean vertical velocities do not show clear variations of angle of line-of-node with $R$. This might largely be due to i) the relative large uncertainties of the mean vertical velocities (see Fig.\,3 and panel (b) of Fig.\,4); and ii) the limited azimuthal angle coverage of the data. In the near future, this issue could be solved by adding more RC stars to the sample, selected from new LAMOST observations and the planned SDSS V surveys. Moreover, the additional data could allow one to explore the dynamical evolution of the Galactic warp (Poggio et al. 2020).

\section{Summary}
In this paper, using a sample of nearly 140,000 RCs with accurate 3D position and 3D velocity measurements, constructed with data from the LAMOST and Gaia surveys, we have explored the kinematic warp signature of the Galactic disk(s). With cuts in the [Fe/H]-[$\alpha$/Fe] plane, 94,028 and 5,212 chemically thin and thick disk stars of $|Z| < 1$\,kpc are selected from the sample. Kinematic signature of warp is clearly detected in the data for the chemically thin disk population, but the signal is not significant for the chemically thick disk population. For the thin disk population, a clear positive gradient of mean vertical velocity as a function of $R$ is found for $R$ between 8 and 13\,kpc. The trend agrees with the recent results from the Gaia DR2 and is also consistent with the prediction of the long-lived large-scale Galactic warp model. The warp signature for the thick disk population is much weaker, largely due to the hot nature of orbits of thick disk stars. For the thin disk stars, we further explore the variations of mean vertical velocity (as a function of $R$) for the different azimuthal slices and find the amplitude of warp increases with $\phi$ and reaches a maximum in slice $\phi \in$ [10, 17] deg. To quantitively determine the angle of line-of-node of the warp, we fit the distribution of mean vertical velocities of the thin disk stars with a long-lived static warp model and find an angle around 12.5$^{\circ}$, in excellent agreement with the previous estimates from star counting.

Based on the current study alone, it is still difficult to constrain the exact origin of the Galactic warp. However, with more data expected from the on-going and forthcoming LAMOST, SDSS and Gaia surveys, vital clues about the origin and evolution of the Galactic warp should become available in the near future.

\section*{Acknowledgements} 
The Guoshoujing Telescope (the Large Sky Area Multi-Object Fiber Spectroscopic Telescope, LAMOST) is a National Major Scientific Project built by the Chinese Academy of Sciences. Funding for the project has been provided by the National Development and Reform Commission. LAMOST is operated and managed by the National Astronomical Observatories, Chinese Academy of Sciences. The LAMOST FELLOWSHIP is supported by Special fund for Advanced Users, budgeted and administrated by Center for Astronomical Mega-Science, Chinese Academy of Sciences (CAMS). R.S. is supported by a Royal Society University Research of Fellowship.

This work has made use of data from the European Space Agency (ESA) mission Gaia (https://www.cosmos.esa.int/gaia), processed by the Gaia Data Processing and Analysis Consortium (DPAC, https://www.cosmos.esa.int/web/gaia/dpac/consortium).

{This work is supported by National Natural Science Foundation of China grants 11903027, 11973001, 11833006, 11811530289, U1731108, and U1531244, and  National Key R \& D Program of China No. 2019YFA0405503. B.Q.C. and Y.H. are supported by the Yunnan University grant No. C176220100006 and C176220100007, respectively. HFW is supported by the LAMOST Fellow project and funded by China Postdoctoral Science Foundation via grant 2019M653504, Yunnan province postdoctoral Directed culture Foundation and the Cultivation Project for LAMOST Scientific Payoff and Research Achievement of CAMS-CAS.}

%\bibliographystyle{apj}
%\bibliography{warp_b}

\end{document}